\documentstyle[epsf]{elsart}
\begin{document}
\begin{frontmatter}
\title  {How to count nucleon pairs?}
\author  {J. Dobe\v{s} \thanksref{rf}}
\address  {   Institute of Nuclear Physics,
   Academy of Sciences of the Czech Republic, \\
     CS 250 68 \v{R}e\v{z}, Czech Republic}
\thanks[rf]{E-mail: dobes@ujf.cas.cz} 
\begin{abstract}
Within the context of the isovector-pairing SO(5) model, three methods
measuring numbers of different kinds of nucleon pairs are discussed.
Though methods do not give the same results, the odd-even staggering in
pair numbers in even-even and odd-odd $N=Z$ nuclei and the reduction of
the np-pair number with increasing $T_z$ is observed in all three
procedures. \\
\end{abstract}
\begin{keyword} Neutron-proton pairing; SO(5) model
\begin{PACS}21.10.-k; 21.60.Fw 
\end{PACS}
\end{keyword}
\end{frontmatter}

Structures comprised of nucleon pairs are influential in describing
nuclear states. The most important pairing degree of freedom is the
monopole pair with $L=0$ orbital angular momentum. In medium-heavy and
heavy nuclei, the monopole pairing pertains to pairs of like nucleons
as valence protons and neutrons occupy different shells and cannot
couple to $L=0$.  Today, considerable experimental activity is aimed at
study of proton-rich nuclei.  In those, the neutron-proton-pairing mode
becomes notable.

In a recent paper \cite{ELV}, a method has been suggested to quantify
effects of different pairing modes and measure pair numbers in the
nuclear-wave function. We should emphasize that the pair numbers are
not directly observable quantities.  Even in simple models of pairing,
there is no operator that counts nucleon pairs. If an exact solution of
the problem is available, one need not to operate with the pair
numbers. Their quantification can, however, make the problem more
transparent. Also in complex situation, the notion of the significance
of different pairs may be helpful in constructing approximate solutions
and/or devising model descriptions.

In the present study, we examine the measure of pair numbers of Ref.
\cite{ELV} and suggest alternative procedures.  As in Ref.
\cite{ELV}, we discuss the problem within the context of a simple model
of monopole-isovector pairing with the underlying O(5) algebraic
structure
\cite{SO5}. The model reflects pairing features of realistic shell-model 
calculations \cite{LD}.

In the monopole-isovector-pairing model, the set of pair creation
operators 
\begin{equation}
S^{\dagger}_{\nu}=\frac{1}{\sqrt{2}}
\sum_{j} [a^{\dagger j \frac{1}{2}} 
a^{\dagger j \frac{1}{2}}]^{01}_{0\nu}
\; \; ,
\end{equation}
the conjugate pair annihilation operators $S_{\nu}$, the isospin
operator $T$, and the total number operator $\hat{N}$ closes the O(5)
algebra. The degeneracy of the shell is $4\Omega = 2\sum (2j+1)$.  The
problem of the SO(5) pairing Hamiltonian
\begin{equation}
H = -G \sum_{\nu} S^{\dagger}_{\nu} S_{\nu} \; ,
\label{hamilt}
\end{equation}
can be solved algebraically. We shall consider the seniority zero
eigenstates $|{\cal N} T T_z \rangle$ of $H$ labelled by half 
the number of nucleons ${\cal N}$,
isospin $T$, and isospin $z$-projection $T_z$.

The basic building blocks of the SO(5) model space are the nn-, pp-, and
np-pair creation operators. We shall examine possible procedures which
would estimate the average numbers $\langle {\cal N}_{{\rm nn}}
\rangle$,
$\langle {\cal N}_{{\rm pp}} \rangle$, 
and $\langle {\cal N}_{{\rm np}} \rangle$ 
of  the respective pairs in a given state.

Within the O(5) algebra, there are no operators which could be
considered as the number operators of particular pairs.  There are,
however, two operators which relate the different pair numbers. Namely,
the sum of the pair numbers should give half the number of nucleons
\begin{equation}
 \langle {\cal N}_{{\rm nn}} \rangle +
 \langle {\cal N}_{{\rm pp}} \rangle + 
 \langle {\cal N}_{{\rm np}} \rangle = {\cal N}
\label{sumpair}
\end{equation}
and the difference between nn- and pp-pair numbers should link with the
charge of the state
\begin{equation}
 \langle {\cal N}_{nn} \rangle -
 \langle {\cal N}_{pp} \rangle =T_z   \; \; .
\label{charge}
\end{equation}

In Ref.  \cite{ELV}, the operator
\begin{equation}
{\cal N}_{{\rm nn}}^{\rm I}=\frac{1}{\Omega} S^{\dagger}_{1} S_{1}
\label{defI}
\end{equation}
and the analogously defined operators ${\cal N}_{{\rm pp}}^{\rm I}$ and
${\cal N}_{{\rm nn}}^{\rm I}$ have been suggested as measures of the
numbers of different pairs.  We denote this method by I.  In this
method, the pair numbers are associated with the contributions of the
respective pairing terms in the Hamiltonian (\ref{hamilt}) to the total
energy and the relations can be obtained
\begin{equation}
 \langle {\cal N}_{{\rm nn}}^{\rm I} \rangle +
 \langle {\cal N}_{{\rm pp}}^{\rm I} \rangle + 
 \langle {\cal N}_{{\rm np}}^{\rm I} \rangle = {\cal N}
(1-\frac{{\cal N}-3}{2\Omega})-\frac{1}{2\Omega} T(T+1)
\label{sumI}
\end{equation}
\begin{equation}
 \langle {\cal N}_{{\rm nn}}^{\rm I} \rangle -
 \langle {\cal N}_{{\rm pp}}^{\rm I} \rangle =
T_z (1-\frac{{\cal N}-1}{\Omega})
\label{chargeI}
\end{equation}
\begin{eqnarray}
\langle {\cal N}_{{\rm np}}^{\rm I} \rangle  &=&
\frac{1}{3}{\cal N}(1-\frac{{\cal N}-3}{2\Omega})
-\frac{1}{6\Omega}T(T+1)
\nonumber \\
& &
-\frac{1}{3} \frac{3T_{z}^2-T(T+1)}{(2T-1)(2T+3)}
[(1-\frac{{\cal N}-1}{\Omega})(2{\cal N}+3)
\nonumber \\
& &
+\frac{1}{\Omega} ({\cal N}+T+1)({\cal N}-T)]
\; .
\end{eqnarray}

For the large shell degeneracy $\Omega$ or more strictly speaking for
${\cal N} \ll \Omega$, the pair operators exhibit boson-like behaviour
and the pair numbers in method I obey Eqs. (\ref{sumpair}) and
(\ref{charge}). For nuclei around the middle of the shell, however,
serious discrepancy between (\ref{sumI}) and (\ref{chargeI}) and the
natural constraints (\ref{sumpair}) and (\ref{charge}) on pair numbers
may occur. One can remedy the total pair number condition by scaling
simply pair numbers in method I (and we shall use this procedure in the
following). Anyhow, the charge constraint (\ref{charge}) will not be
fulfilled.

This deficiency puts some question on the method I as a measure of
different pair numbers. Certainly, operators (\ref{defI}) quantify
contributions of different pairing terms in Hamiltonian (\ref{hamilt}).
In many-pair states, however, those are influenced by the Pauli
principle and the relation to the pair numbers is not immediate and
even may be misleading.

We shall propose an alternative method II for obtaining the pair
numbers that employs directly the pair basis. In that the states are
constructed as
\begin{equation}
|{\cal N}_{{\rm nn}} {\cal N}_{{\rm pp}} {\cal N}_{{\rm np}} \rangle =
a S^{\dagger {\cal N}_{{\rm nn}}}_{1}
S^{\dagger {\cal N}_{{\rm pp}}}_{-1}
S^{\dagger {\cal N}_{{\rm np}}}_{0} |0 \rangle
\; ,
\label{pairbasis}
\end{equation}
where $a$ normalizes the state (\ref{pairbasis}).  The use of the pair
basis in calculating the average pair numbers is not straightforward.
Namely, the pair basis is not orthogonal and for ${\cal N} > \Omega$ is
even overcomplete.  Nevertheless, the squared overlap of the pair state
(\ref{pairbasis}) with a wave function $|\psi
\rangle $ gives certainly a probability of the occurrence
of the state with given pair numbers.  We are thus tempted  to define
the average pair numbers as 
\begin{equation}
 \langle {\cal N}_{{\rm nn}}^{\rm II} \rangle =
 C \sum {\cal N}_{{\rm nn}} | \langle {\cal N}_{{\rm nn}} 
{\cal N}_{{\rm pp}} {\cal N}_{{\rm np}} |
\psi \rangle |^2
\label{defII}
\end{equation}
with the scaling factor
\[
C  = \sum | \langle {\cal N}_{{\rm nn}} {\cal N}_{{\rm pp}} 
{\cal N}_{{\rm np}} |
\psi \rangle |^2
\; 
\]
and similarly for $\langle {\cal N}_{{\rm pp}}^{\rm II} \rangle$ and
$\langle {\cal N}_{{\rm np}}^{\rm II} \rangle$.  Note that $C$ scales
the definition (\ref{defII}) so that Eqs. (\ref{sumpair}) and
(\ref{charge}) are obeyed.  In the limit of large $\Omega$, pair
numbers from method II agree with those from method I.

The complications in the preceding two definitions (\ref{defI}) and
(\ref{defII}) of pair numbers are connected with the fact that we
intend to implement the boson-like behaviour to pair operators that do
not obey the boson commutation relations exactly unless $\Omega$ is
large. This mirrors in the insufficiency of the method I to fulfil the
charge constraint (\ref{charge}) and in the necessity to deal with the
nonorthogonal basis in the method II. 

There exists a possibility to rephrase the fermion problem in the boson
language.\footnote {For a detailed review on boson realizations of
fermion Lie algebras see \cite{KM}.} For the bifermion O(5) algebra,
the Dyson boson realization has been constructed in Ref. \cite{GH}.  In
our specific isovector-pairing case, the boson space is formed from the
scalar-isovector boson $s_{\nu}$. The fermion-pair operators are mapped
on the boson operators as
\[ S^{\dagger}_{\nu} \rightarrow
(\Omega - \frac{1}{2} \hat{N}+1)s^{\dagger}_{\nu}
+(-)^{\nu} \frac{1}{2} s^{\dagger} \cdot s^{\dagger}
{s}_{-\nu}
\; ,
\]
\begin{equation}
S_{\nu} \rightarrow s_{\nu}
\; .
\label{Dyson}
\end{equation}

Using the above relations, the boson image $H_{\rm B}$ of the
Hamiltonian $H$ is constructed. When $H_{\rm B}$ is diagonalized in the
ideal (Fock) boson space, all the eigenvalues and boson images of
eigenstates of the original fermion problem are provided.  The
dimension of the ideal boson space can be larger that that of the
fermion task. For the SO(5) Hamiltonian, spurious states occur in the
boson picture for ${\cal N} > \Omega$.  They, however, clearly separate
from the physical solutions for example by the isospin value.\footnote
{The boson mapping may provide a simple technique to solve the
fermion problem. In fact, we use it to obtain the 
pair numbers in both methods I and II .}

\begin{figure}[hbt]
\epsfxsize 10.cm
\centerline{\epsfbox{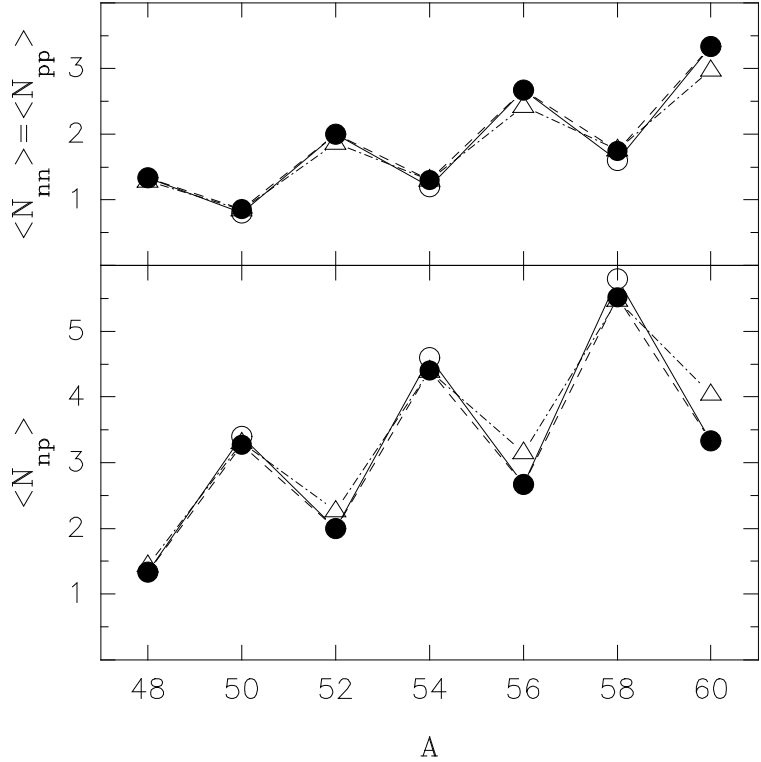}}
\caption{The pair numbers for the SO(5) model ground state 
in $N=Z$ nuclei.
The filled circles, triangles, and hollow circles refer to methods
I, II, and III, respectively.
}
\label{fig1}
\end{figure}

We may make use of the boson formulation of the fermion problem to get
the boson-like fermion-pair numbers.  The bosons are images of the
fermion pairs.  In the boson space, the boson number operators
$\hat{n}$ are well defined.
We identify the average boson numbers with another measure (method III)
of the pair numbers
\begin{equation}
 \langle {\cal N}_{nn}^{\rm III} \rangle =
( \hat{n}_{s_1} )
\label{defIII}
\end{equation} 
and similarly for $ \langle {\cal N}_{pp}^{\rm III} \rangle$ and $
\langle
{\cal N}_{np}^{\rm III} \rangle$.  Here, the boson eigenvectors are 
denoted by the round brackets.\footnote
{Note that the Dyson boson mapping is a nonunitary mapping so that the
boson Hamiltonian may generally be nonhermitian ( this is not the case
for the Hamiltonian (\ref{hamilt})) with different left and right boson
eigenvectors.} Such a definition of pair numbers obeys
Eqs. (\ref{sumpair}) and (\ref{charge}). Explicitly, one gets 
\begin{eqnarray}
\langle {\cal N}_{np}^{\rm III} \rangle  =
\frac{1}{3}{\cal N}
-\frac{1}{3} \frac{3T_{z}^2-T(T+1)}{(2T-1)(2T+3)}
(2{\cal N}+3)
\; .
\end{eqnarray}

\begin{figure}[tb]
\epsfxsize 13.8cm
\centerline{\epsfbox{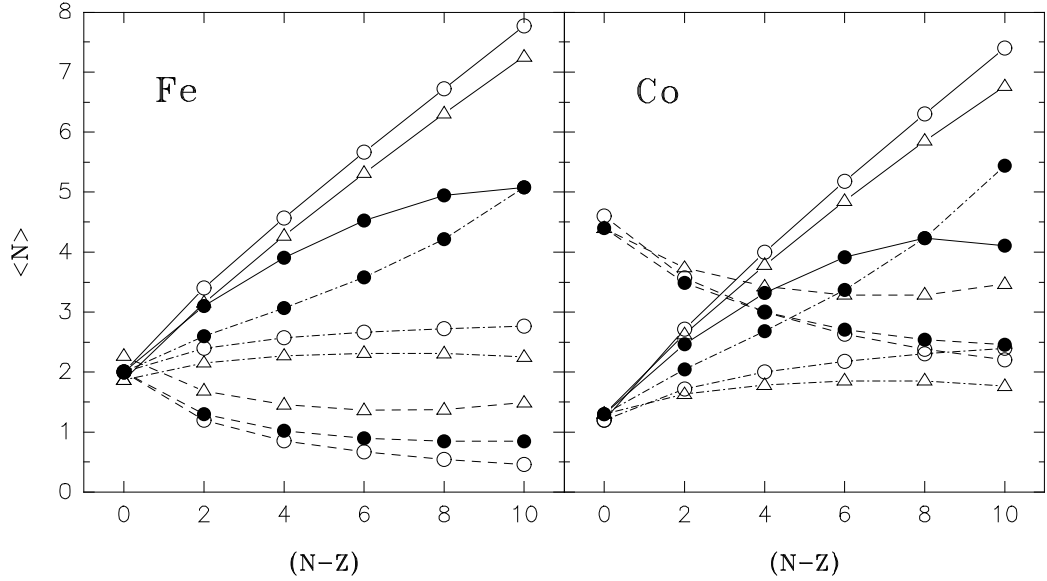}}
\caption{
The pair numbers for the SO(5) model ground state 
in even-even Fe and odd-odd Co nuclei.
The filled circles, triangles, and hollow circles refer to methods
I, II, and III, respectively. 
The full lines connect $\langle {\cal N}_{{\rm nn}} \rangle$ points,
the dash-dotted lines connect 
$\langle {\cal N}_{{\rm pp} \rangle}$ points,
and the dashed lines connect $\langle {\cal N}_{{\rm np} \rangle}$ 
points.
}
\label{fig2}
\end{figure}

Alternatively, the boson definition (\ref{defIII}) of pair numbers is
obtained from the method I employing the pairing force when the
$1/\Omega$ corrections are neglected.\footnote
{In the present SO(5) case, we know the results in method I
analytically and can perform the $\Omega \rightarrow \infty$ limit without
resorting to the boson formulation. In the general case, the
diagonalization in the boson space is necessary.} One cannot, however,
deduce that the definition (\ref{defIII}) is inferior to that of
(\ref{defI}). The $1/\Omega$ corrections reflect the Pauli principle
and deviations of the pair behaviour from the boson one. On the other
hand, we would like to understand the pair numbers which are the
boson-like quantities. Simple relations of Eqs. (\ref{sumpair}) and
(\ref{charge}) for the pair numbers should hold which are in the method
I deteriorated just by the $1/\Omega$ terms.

As it is seen from Fig. \ref{fig1}, all three methods give very similar
quantities for the average pair numbers in the ground states of $N=Z$
$fp$ shell nuclei. The odd-even staggering observed in Ref. \cite{ELV}
is thus confirmed also by another two measures of the numbers of
different pairs.\footnote {Note that the proximity of the results of
the methods I and III does not mean that the $1/\Omega$ corrections in
(\ref{defI}) are small. We remind that the pair numbers from the
definition (\ref{defI}) are scaled to sum to the total number of
pairs.}

In Fig. \ref{fig2} on the left,
 the pair numbers for even-even  Fe isotopes
are shown.  Here, the three methods differ more pronouncedly than for
the case of Fig. \ref{fig1}.  Nevertheless, the alternative methods
confirm a decrease of the number of np pairs with increasing $T_z$
noticed in Ref. \cite{ELV}.  In the method I, numbers of nn and pp
pairs increase with increasing $T_z$ and significantly break down the
charge condition (\ref{charge}).  On the other hand, in methods II and
III that fulfil the charge constraint, a rough constancy of the number
of pp pairs and linear increase of the number of nn pairs is observed
when the neutrons are added to the nucleus. In Fig. \ref{fig2} on the
right,  the
odd-odd  Co isotopes exhibit a similar behaviour with increased
importance of the np pair as compared to even-even case.

When $T_z$ is increasing, the decrease of the number of np pairs 
suggests a diminishing role of the np pairing.
It would be
important to be able to judge at what value of $(N-Z)$ one can neglect
the np-pairing degree of freedom and approximate the ground state as a
product of nn and pp paired states. The exactly solvable SO(5) model 
makes possible such a discussion. The exact ground-state wave function
of the Hamiltonian (\ref{hamilt}) for the even-even nucleus is written
up to the normalization factor as
\begin{equation}
|{\cal N} \, T\!=\!T_z  \,T_z \rangle \sim 
(S^{\dagger} \cdot S^{\dagger})^{({\cal N}-T_z)/2} 
S^{\dagger T_z}_{1} |0 \rangle
\label{evenex}
\end{equation}
whereas for the odd-odd nucleus we have
\begin{equation}
|{\cal N} \, T\!=\!T_z\!+\!1 \, T_z \rangle \sim 
(S^{\dagger} \cdot S^{\dagger})^{({\cal N}-T_z-1)/2} 
S^{\dagger T_z}_{1} S^{\dagger}_{0} |0 \rangle
\; .
\label{oddex}
\end{equation}
The important building block of the ground-state wave function is 
the two-pair $T=0$ structure  
\begin{equation}
S^{\dagger} \cdot S^{\dagger}=
S^{\dagger}_{0}S^{\dagger}_{0}-2S^{\dagger}_{1}S^{\dagger}_{-1}
\; .
\label{twopairT=0}
\end{equation}
From that the $T=0$ even-even core is constructed to which an
appropriate number of nn pairs and one np pair for the odd-odd case
is added. 

We shall approximate the SO(5) ground state 
by discarding the np part in 
the above two-pair $T=0$ structure (\ref{twopairT=0}).
The isospin is then mixed in the approximate solution.
For the even-even nucleus, we have
\begin{equation}
|{\cal N}  T_z  \: , \: {\rm no \; np \; correlations} \rangle \sim 
S^{\dagger({\cal N}+T_z)/2 }_{1}  S^{\dagger({\cal N}-T_z)/2}_{-1} 
|0 \rangle
\; .
\end{equation}
Of course, at least  one np pair must be present for the odd-odd nucleus
\begin{equation}
|{\cal N}  T_z \: , \: {\rm no \; np \; correlations} \rangle \sim 
S^{\dagger ({\cal N}+T_z-1)/2 }_{1}  
S^{\dagger ({\cal N}-T_z-1)/2}_{-1} 
S^{\dagger}_{0} |0 \rangle
\; .
\end{equation}

\begin{figure}[tb]
\epsfxsize 13.8cm
\centerline{\epsfbox{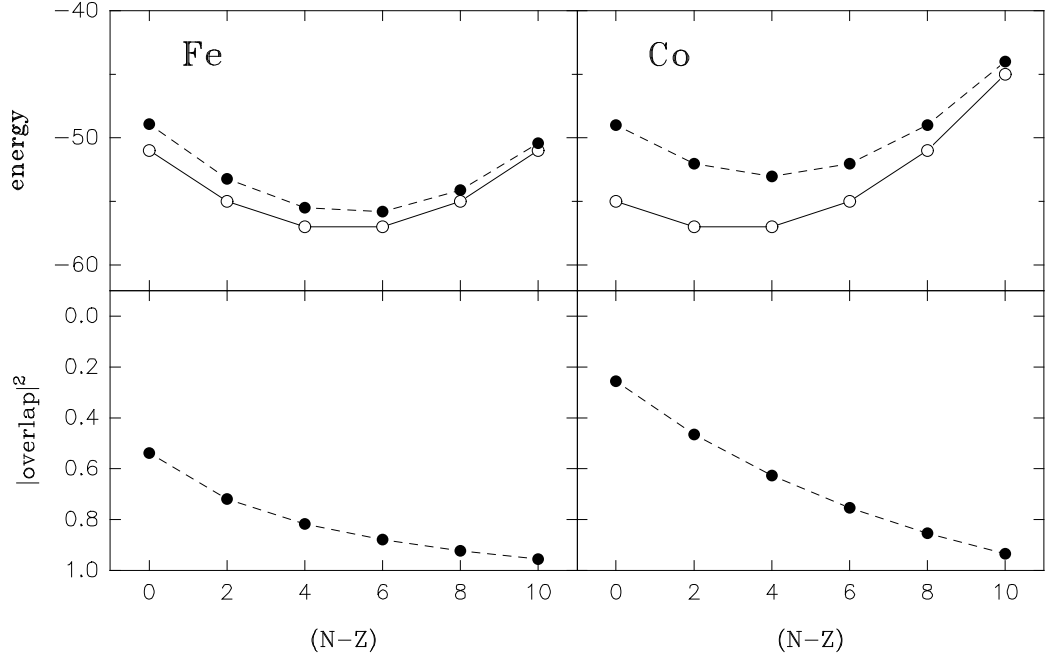}}
\caption{
In the upper part, the exact ground-state energy (in units of $G$)
(hollow circles) and
the mean value of $H$ in the approximate state with no np
correlations (filled circles) are given. In the lower part, 
the squared overlap
of the exact and approximate wave functions is shown.
}
\label{fig3}
\end{figure}

In the upper part of Fig.\ref{fig3}, the exact ground-state energy is
compared with the mean value of $H$ in the approximate state with no np
correlations. In the lower part of Fig.\ref{fig3}, the squared overlap
of the exact and approximate wave functions is shown. 
Fig.\ref{fig3} corresponds to the findings inferred from the analysis of pair
numbers. The diminishing role of the np pair with
increasing $T_z$ is particularly
exhibited in the even-even case.  In the odd-odd nucleus, the
mandatory np pair stabilizes the np-pairing degree of freedom 
and  there is  a little overlap between the exact state and the
wave function with the np correlations neglected for
small values of $T_z$.
Here, however, the increase of the overlap is more abrupt than in the
even-even case. For $(N-Z)=10$ both in the even-even and odd-odd
case, the exact wave function overlaps 
with the
wave function with no np correlations
by more than 90\%.

Note also that the exact and approximate ground-state energies might not 
differ much even if the overlap is small. The leading term
in the ground-state energy is given as $-G\Omega {\cal N}$ 
irrespective
of the actual form of the wave function. The binding energies alone are
not thus best quantities to judge about the np correlations and
about the quality of an approximate wave function. One
should resort to difference filters, such as that of Ref. \cite{ZCB},
in which the leading term contribution is eliminated.

If the isospin symmetry is violated in the SO(5) Hamiltonian and the 
np-pairing force is set to zero, 
one would expect that the number of np
pairs should go to zero. However, the methods I and II give a nonzero
average number of np pairs in this case.  Similarly, in the O(8) space
including the isoscalar pair \cite{EPSVD}, the number of isoscalar pairs
is nonzero for the Hamiltonian with the isoscalar pairing switched off
in methods analogous to (\ref{defI}) and (\ref{defII}). On the other
hand, the boson method III of counting pairs gives a null number for
the pairing mode whose pairing term is not present in the Hamiltonian.

To conclude, we stress again that counting the number of pairs in a
fermion state is not well and completely determined task. There is no
fermion operator which could be identified with the number operator of
a particular pair.  Nevertheless, three possibilities have been
investigated in the present paper how to get quantities that may be
related to the pair numbers.  Though methods give different results,
two important conclusions of Ref. \cite{ELV} remain unchanged. Namely,
the odd-even staggering in pair numbers in even-even and odd-odd $N=Z$
nuclei and the reduction of the np-pair number with increasing $T_z$ are
observed in all three procedures. The boson counting method seems to
have certain advantages as it obeys the natural relations
(\ref{sumpair}) and
(\ref{charge}) for pair numbers
 and gives a zero pair number for the particular pairing
mode switched off in the Hamiltonian.
\\

We thank J. Dobaczewski, H.B. Geyer, and P. Vogel for useful and
stimulating discussions. This work has been supported by the Grant
Agency of the Academy of Sciences of the Czech Republic under grant No.
A1048504.


\end{document}